\documentclass[aps,prd,twocolumn,preprintnumbers,nofootinbib]{revtex4-1}
\usepackage{amsmath, amssymb, graphicx, setspace,indentfirst,gensymb,color,slashed,url}

\usepackage{tikz-feynman}
\tikzfeynmanset{compat=1.0.0}

\newcommand{\nc}{\newcommand}
\nc{\Tr}{\mbox{Tr}}
\nc{\hc}{\mbox{H.c.}}
\nc{\ev}{\;\mathrm{eV}}
\nc{\mev}{\;\mathrm{MeV}}
\nc{\gev}{\;\mathrm{GeV}}
\nc{\infinity}{\infty}

\newcommand{\<}{\langle}
\renewcommand{\>}{\rangle}
\newcommand{\eee}{\equiv}
\renewcommand{\it}[1]{\textit{#1}}

\nc{\beq}{\begin{equation}}
\nc{\eeq}{\end{equation}}

\begin{document}
\title{Faint Dark Matter Annihilation Signals and the Milky Way's Supermassive Black Hole}
\author{Barry T. Chiang}
\email{tchiang5@illinois.edu}

\author{Stuart L. Shapiro}
\email{slshapir@illinois.edu}

\author{Jessie Shelton}
\email{sheltonj@illinois.edu}

\affiliation{\textit{Departments of Physics and Astronomy, University of Illinois at
	Urbana-Champaign, Urbana, IL 61801, USA}}
\begin{abstract}
A wide range of mechanisms predict present-day $s$-wave dark matter (DM) annihilation cross-sections that are orders of magnitude below current experimental sensitivity.
We explore the capability of DM density spikes around the Milky Way's supermassive black hole to probe such faint signals of DM annihilations, considering a range of possible spike and halo distributions. 
As an exemplar of a theory with a suppressed $s$-wave annihilation cross-section, we consider a hidden sector axion portal model of DM.  In this model, the leading contribution to the annihilation cross-section in the early universe is $p$-wave, while $s$-wave annihilations occur at higher order in the coupling constant.   We provide a unified treatment of DM freezeout in this model including both $s$- and $p$-wave annihilations and analytically determine the photon spectrum for the dominant DM annihilation process in the universe today.  We find that \textit{Fermi} and H.E.S.S. observations of the Galactic Center offer excellent sensitivity to this model over a wide range of parameter space, with prospects depending sensitively on the properties of the DM spike as well as the central halo.
\end{abstract}
\maketitle

	\section{Introduction}

The indirect detection of dark matter through its annihilation products in cosmic rays is a cornerstone of the experimental search for dark matter (DM). Indirect detection is an increasingly potent probe of annihilating DM, with observations of (e.g.)  both the cosmic microwave background and dwarf galaxies now sensitive to DM with annihilation rates at or below the standard thermal target $\sigma_\text{thermal}= 3\times 10^{-26}\,\mathrm{cm}^3/$s in an expanding range of masses and annihilation channels  \cite{Ade:2015xua,Slatyer:2015jla,Ackermann:2015zua,Fermi-LAT:2016uux}.  

Identifying the annihilation products of TeV-scale DM with standard thermal-scale cross-sections remains a steep observational challenge, however, as the flux of cosmic rays from DM annihilation in galaxy halos falls off with increasing DM mass as $m^{-2}$.  
Moreover, many dark matter models predict present-day DM annihilation cross-sections substantially below the thermal target. There are many mechanisms that predict a suppressed present-day DM annihilation cross-section. For instance, the DM may simply arise from a dark sector that is very cold compared to the Standard Model (SM) \cite{Feng:2008mu,Evans:2019vxr}; it may have been diluted by entropy production post-freezeout \cite{Fornengo:2002db, Gelmini:2006pq, Hooper:2013nia}; its annihilation cross-section may be strongly velocity-dependent, because of kinematics \cite{Griest:1990kh,DAgnolo:2015ujb,Kopp:2016yji,Delgado:2016umt} or symmetries \cite{Goldberg:1983nd,Griest:1988ma,Boehm:2003hm,Shelton:2015aqa,Evans:2017kti} (or both); or its annihilation cross-section in the early universe may involve additional species that are later depleted \cite{Griest:1990kh,Cirelli:2011ac,Baker:2015qna,Garny:2017rxs,DAgnolo:2017dbv}. In several of these models, e.g.~\cite{Feng:2008mu,Shelton:2015aqa,Evans:2019vxr}, terrestrial signals are typically significantly reduced compared to expectations from thermal WIMPs, making even a suppressed indirect detection signal an irreplaceable discovery handle and a powerful window onto the physics of DM.

Here we estimate the sensitivity to faint $s$-wave DM annihilation cross-sections that can potentially be offered by DM density spikes around the supermassive black hole (SMBH) at the center of the Milky Way. Black holes focus DM within their gravitational zone of influence into a steep, localized over-density known as a spike \cite{Gondolo:1999ef}. This enhancement of the DM density can potentially magnify DM annihilation rates by many orders of magnitude in the immediate vicinity of the black hole, leading to a bright, point-like source of cosmic rays. While this enormous magnification of DM annihilation signals can offer a uniquely powerful window onto models of DM with suppressed annihilation cross-sections and thus suppressed signatures in DM haloes \cite{Amin:2007ir,Cannoni:2012rv,Lacroix:2015lxa,Shelton:2015aqa, Johnson:2019hsm}, the details of the predicted spike distribution depend sensitively on as-yet-unknown properties of the host DM halo, the central stellar cusp within that halo, and the formation history of the black hole  \cite{Gondolo:1999ef,Nakano:1999ta,Ullio:2001fb,Merritt:2003qk, Gnedin:2003rj, Merritt:2006mt,Shapiro:2014oha}. Accordingly,  we consider a broad range of possible DM distributions in the Galactic Center in this work, with the aim of understanding what parameter ranges offer interesting sensitivity to sub-thermal annihilation cross-sections.   Uncertainties in the spike distribution translate into very large uncertainties in possible DM signal strengths, making it hard to unambiguously constrain DM models using annihilation signals within DM spikes.  Nonetheless DM spikes can provide an invaluable potential opportunity for {\em discovery}, particularly for models with suppressed annihilation cross-sections and sharp spectral features.

As a representative DM model with both challenging annihilation cross-sections and sharp spectral features, we consider the hidden sector axion portal (HSAP) model developed in \cite{Shelton:2015aqa, Johnson:2019hsm}. In this model, fermionic DM $\chi$ annihilates to pseudo-scalars $a$, which subsequently decay to SM electroweak gauge bosons. This model features an interesting interplay of two annihilation channels: the process $\chi\chi \to a a$ proceeds in the $p$-wave, while the reaction $\chi\chi\to aaa $ contributes in the $s$-wave when it is kinematically available, but is higher order in the coupling constant.  For low DM masses, the $p$-wave process dominates DM annihilation in the early universe, yielding a subdominant $s$-wave annihilation cross-section orders of magnitude below the standard thermal target. For high DM masses, the $s$-wave annihilation process can dominate during thermal freezeout, resulting in thermal-scale cross-sections but signals that are observationally challenging thanks to the large DM mass.  In all cases the $s$-wave process $\chi\chi\to aaa $ dominates the present-day DM annihilation rate, including within DM spikes.   

We begin in Section~\ref{sec:hsap} with an overview of the HSAP model, including a novel treatment of freezeout incorporating both $s$- and $p$-wave contributions to the annihilation cross-section, and a calculation of the photon spectrum from the resulting DM annihilations. In Sec.~\ref{sec:spike} we detail our model of DM density spikes around the Milky Way's SMBH. We compare predicted gamma-ray fluxes from DM annihilation within SMBH-induced density spikes to observations from \textit{Fermi} and H.E.S.S. in Sec.~\ref{sec:data} and discuss the resulting prospects for sensitivity, and in Sec.~\ref{sec:conclude} we conclude.

\section{The Particle Model}
\label{sec:hsap}

In this section we discuss a hidden sector axion portal model of dark matter, as introduced in \cite{Shelton:2015aqa}. 
In this model, DM is a Majorana fermion $\chi$ which annihilates to pseudoscalars $a$, which subsequently decay to the SM via axion-like couplings to SM gauge bosons. This model is $CP$-conserving, ensuring that the leading annihilation process $\chi\chi\to aa$ is $p$-wave.
In contrast to previous works \cite{Shelton:2015aqa, Johnson:2019hsm}, we will focus on the regime where the higher-order but $s$-wave annihilation process $\chi\chi\to aaa$ is kinematically available, and explore the consequences both for thermal freezeout and for potential BH spike signals in our Galaxy today. 

\subsection{Annihilations and Relic Abundance}

The hidden sector axion portal (HSAP) model is described by the Lagrangian
\begin{equation}
  \mathcal{L} =  \bar\chi (i\gamma\cdot\partial - m_\chi )\chi  -\frac{1}{2}(\partial a)^2  -\frac{1}{2} m_a^2 a^2 - i y 
  a\,  \bar \chi \gamma^5 \chi .
\end{equation} 
This Lagrangian has three free parameters: the masses $m_\chi$ and $m_a$, and the Yukawa coupling constant $y$, which can be determined in terms of $m_\chi$ and $m_a$ using the requirement that thermal freezeout of DM annihilations yields the observed DM relic abundance.  Additionally, the mediator $a$ is coupled to the SM via dimension-five axion-like interactions with SM gauge fields, which enables it to decay promptly on astrophysical scales, as we discuss further below.

\subsubsection{Annihilation cross-sections}

The leading $2\to 2$ annihilation process occurs in the $p$-wave, with thermally averaged cross-section given in the non-relativistic limit by \cite{Shelton:2015aqa}
\begin{gather}
\< \sigma v\>_p = \frac{1}{x}\frac{y^4}{4\pi m_\chi^2}\sqrt{1-\eta^2}\frac{(1-\eta^2)^2}{(2-\eta^2)^4},
\label{eqn:pCrossSection}
\end{gather}
where 
\begin{gather}
\eta \eee m_a/m_\chi
\label{eqn:MaOverMChi}
\end{gather}
and
\begin{gather}
x \eee m_\chi/T.
\label{eqn:xDefinitionFirst}
\end{gather}
The velocity dispersion is related to the temperature as $\< v^2\>=6/x$.

\begin{figure}
\begin{tikzpicture}
\begin{feynman}
\vertex (LeftEdge);
\vertex [right=1.35cm of LeftEdge](bMiddle);
\vertex [above=0.67cm of bMiddle] (b1);
\vertex [below=0.67cm of bMiddle] (b2);
\vertex [above=0.3cm of b1] (A1);
\vertex [below=0.3cm of b2] (A2);
\vertex [left=1.2cm of A1] (a1){\(\chi\)};
\vertex [left =1.2cm of A2] (a2){\(\chi\)};
\vertex [right=1.2cm of A1] (c1){\(a\)};
\vertex [right =1.2cm of A2] (c2){\(a\)};
\vertex (MiddleEdge);
\vertex [right=4.45 cm of bMiddle] (eMiddle);
\vertex [above=0.67cm of eMiddle] (e1);
\vertex [below=0.67cm of eMiddle] (e2);
\vertex [above=0.3cm of e1] (D1);
\vertex [below=0.3cm of e2] (D2);
\vertex [left=1.2cm of D1] (d1){\(\chi\)};
\vertex [left =1.2cm of D2] (d2){\(\chi\)};
\vertex [right=1.2cm of D1] (f1){\(a\)};
\vertex [right =1.2cm of D2] (f2){\(a\)};
\vertex [right =1.2cm of eMiddle] (f3){\(a\)};
\diagram* {
	LeftEdge -- [opacity=0] bMiddle,
(a1) -- [fermion, small] (b1),
(b2) -- [fermion, small] (b1),
(b2) -- [fermion, small] (a2),
(b1) -- [scalar] (c1),
(b2) -- [scalar] (c2),
(d1) -- [fermion, small] (e1),
(e2) -- [fermion, small] (e1),
(e2) -- [fermion, small] (d2),
(e1) -- [scalar] (f1),
(e2) -- [scalar] (f2),
(eMiddle) -- [scalar] (f3),
};
\end{feynman}
\end{tikzpicture}
	\caption{Representative Feynman diagrams for \textit{p}-wave (left) and \textit{s}-wave (right) DM annihilation processes.}
	\label{fig:FeynmanDiagram(LaTeX)}
\end{figure}
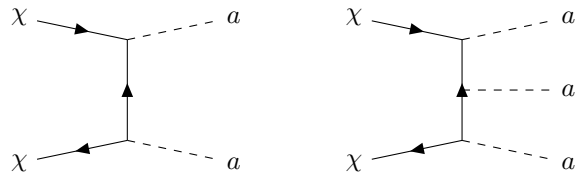

There is one Feynman diagram topology for the process $\chi\chi\to aaa$, shown in Fig.~\ref{fig:FeynmanDiagram(LaTeX)}, which yields six distinct  diagrams thanks to the permutation of final state momenta among the identical particles. The calculation of the matrix element simplifies significantly in the nonrelativistic limit,
where all of the angular dependence drops out. The resulting
expression for the spin-averaged matrix element is
\begin{equation}
| \bar {\mathcal{M}} |^2 = \frac{16 y^6}{m_\chi^2}\, f(w_1,w_2;\eta)
\label{eqn:MatrixElements}
\end{equation} 
where the variables $w_i$ parametrize the distribution of energy among the three final state particles in the CM frame,
\begin{equation}
w_i \equiv \frac{E_i}{E_\chi}, \phantom{space} \sum_{i=1}^3 w_i = 2,
\label{eqn:KinematicsDefinition}
\end{equation}
and
\begin{equation}
f (w_1, w_2; \eta) \equiv \left[ \frac{(1-\frac{\eta^2}{4})(1+\frac{3\eta^2}{4}) + \sum\limits_{i=1}^{3} \left(\frac{1}{2}w_i^2 - w_i\right)}{\prod\limits_{i=1}^{3}\left(w_i - \frac{\eta^2}{2}\right)}\right]^2.
\label{eqn:PhaseSpace}
\end{equation}
In the non-relativistic limit, the cross-section for $\chi\chi\to aaa$ is then 
\begin{equation}
\sigma v = \frac{y^6}{ 96\pi^3 m_\chi^2}\int
 dw_1 dw_2 f(w_1,
w_2;\eta) + \mathcal{O}(v^2), 
\label{eqn:sCrossSection}
\end{equation}
where the upper and lower limits of integration for $w_2$ are given by
\begin{eqnarray}
\nonumber
w_{2\pm} &=& 1- \frac{\eta^2}{4} -2 \varepsilon_j \varepsilon_k \pm \frac{1}{2}\sqrt{(4\varepsilon_j^2-\eta^2)(4\varepsilon_k^2-\eta^2)}, \\
\nonumber
\varepsilon_j &=& \frac{1}{2} \sqrt{1-w_1+\frac{\eta^2}{4}},\\
\varepsilon_k &=& \frac{1}{2} \frac{w_1-\frac{\eta^2}{2}}{\sqrt{1-w_1+\frac{\eta^2}{4}}} \, ,
\label{eqn:w2Limits}
\end{eqnarray}
and for $w_1$ the upper and lower limits are
\beq
w_{1-} = \eta, \phantom{sp} w_{1+} = 1-\frac{3}{4}\eta^2.
\label{eqn:w1Limits}
\eeq
The velocity-dependent $\mathcal{O}(v^2)$ term is negligible both at freezeout (where it is higher order in $y^2/4\pi$ compared to the contribution from $\chi\chi\to aa$) and in the Milky Way today. 
We thus retain only the  piece of Eq.~\ref{eqn:sCrossSection} that is constant as $\langle v^2\rangle \to 0$, defining a contribution to the $s$-wave annihilation $\< \sigma v\>_s$.

\subsubsection{Thermal freezeout and relic abundance}

To determine the Yukawa coupling constant $y$, we include the leading contributions to both \textit{s}- and \textit{p}-wave annihilation processes and solve the Boltzmann equation governing the DM relic abundance. In the non-relativistic limit, this Boltzmann equation 
can be written as
\begin{gather}
\frac{dY}{dx} = -\lambda x^{-2}
(\< \sigma v\>_s+\< \sigma v\>_p)(Y^2-Y^2_{\text{eq}}).
\label{eqn:Boltzmann equation}
\end{gather}
where $\lambda = 0.264 (g_{*S}/ g_*^{1/2}) m_\text{Pl} m_\chi$ and the equilibrium yield is $Y_\text{eq} = 0.29(g_{*S})^{-1}x^{3/2} e^{-x}$ \cite{Kolb:1990}. Here $m_\text{Pl} = G^{-1/2}$ is the Planck mass, and $g_{*}$ and $g_{*S}$ are the number of effective relativistic degrees of freedom contributing to the energy and entropy densities, respectively. The present-day DM relic abundance is taken to be $\Omega_\text{DM} h^2 = 0.112$ \cite{Ade:2015xua}.
%

\begin{figure}
	\includegraphics[scale=0.178]{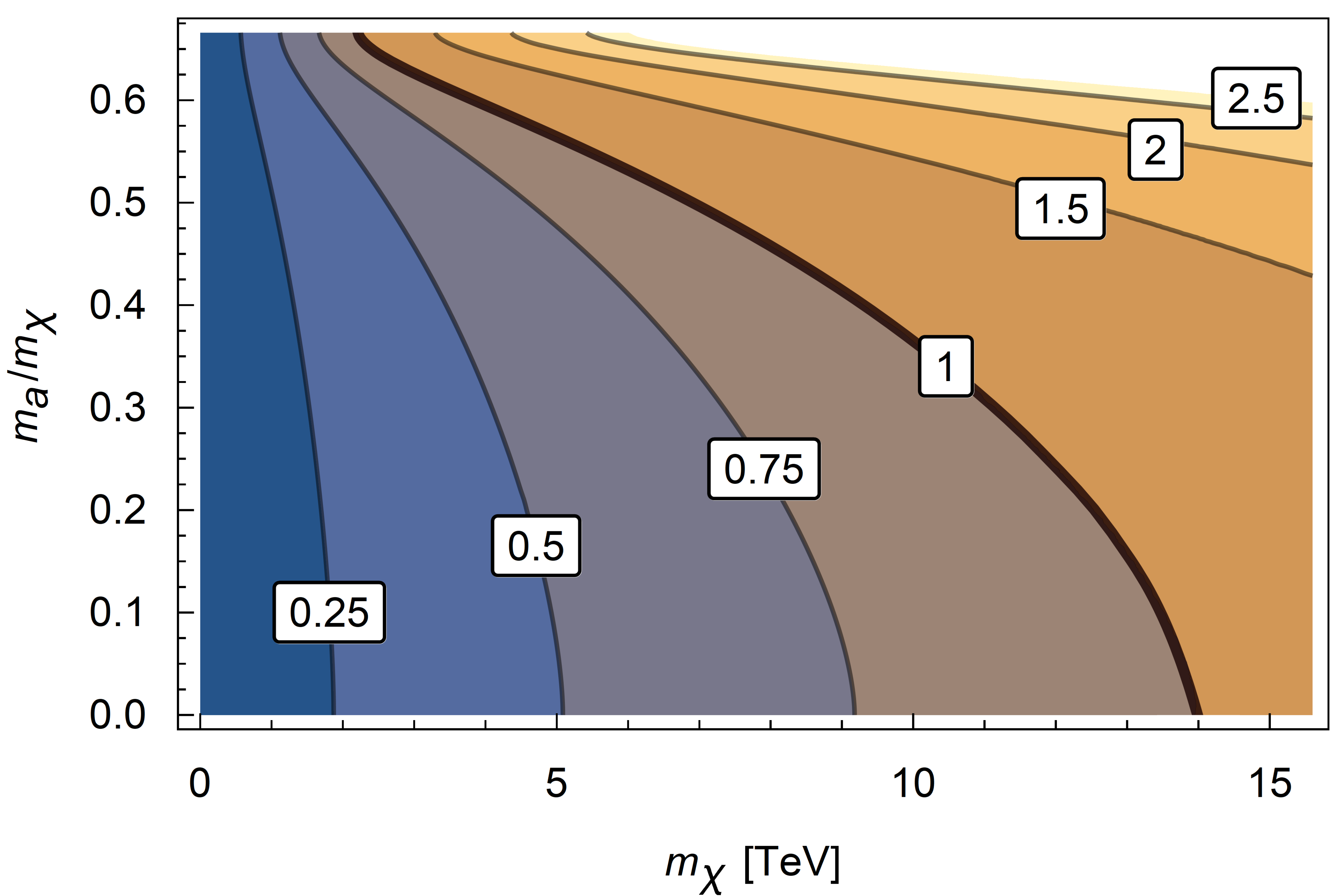}
	\caption{Value of $y^2/4\pi$ yielding the correct DM relic abundance as a function of $m_\chi$ and $\eta = m_a/m_\chi$. The thick black contour indicates $y^2/4\pi = 1$, beyond which the model becomes non-perturbative.}
	\label{fig:y^2over4pi}
\end{figure}

Figure~\ref{fig:y^2over4pi} shows the resulting contours of $y^2/4\pi$ in the $m_\chi$\textendash$\eta$ parameter space. The model becomes non-perturbative for $y^2/4\pi \gtrsim 1$, which restricts $m_\chi\lesssim 14$ TeV. The interplay between \textit{s}- and \textit{p}-wave contributions during freezeout becomes especially important for heavier $m_\chi$ and smaller $\eta$, where the $\chi\chi\to aaa$ annihilation is less suppressed compared to the $\chi\chi\to aa$ process.

\begin{figure}
	\includegraphics[scale=0.66]{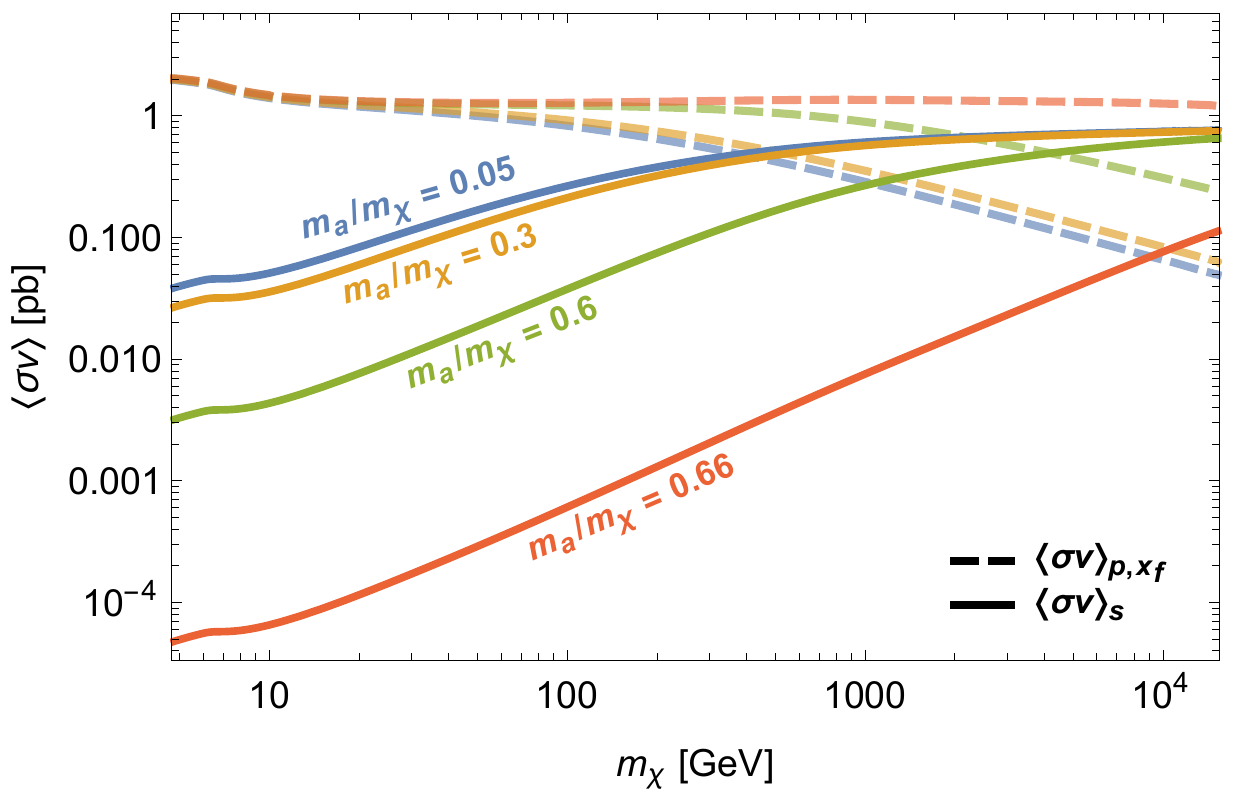}
	\caption{Thermal \textit{s}-wave (solid) and \textit{p}-wave (dashed) annihilation cross-sections, evaluated at $x_f$, for $\eta = 0.05$ (blue), $0.3$ (yellow), $0.6$ (green), and $0.66$ (red). The \textit{s}-wave annihilation contribution becomes appreciable at thermal freezeout for large $m_\chi$.}
	\label{fig:CrossSection}
\end{figure}

Figure~\ref{fig:CrossSection} compares the \textit{s}-wave (solid) and \textit{p}-wave (dashed) annihilation cross-sections for four representative values of the mass ratio, $\eta = 0.05$ (blue), $0.3$ (yellow), $0.6$ (green), and $0.66$ (red). In this figure we evaluate the $p$-wave annihilation cross-section at freezeout, $x_f$, as determined through the sudden freezeout condition
\begin{gather}
n (x_f)\left(\<\sigma v\>_s+\<\sigma v\>_p (x_f)\right) \approx H (x_f) .
\label{eqn:Freezeout}
\end{gather}
Here the Hubble rate is given by $H(x_f) = 1.66 \sqrt{g_*} m_\chi^2/m_\text{Pl}\, x_f^{-2}$.
The resulting values of $x_f$ range from $22<x_f<32$ for $5\,\mathrm{GeV}<m_\chi <14$~TeV and $0<\eta <2/3$.  Since $x_f$ increases with increasing $m_\chi$, at large $m_\chi$ the $s$-wave cross-section  becomes increasingly important at $x_f$ compared to the velocity-suppressed $p$-wave contribution, and can dominate freezeout for sufficiently heavy DM and sufficiently small $\eta$.  The size of the $s$-wave cross-section depends sensitively on the mass ratio $\eta$, and is strongly suppressed as $\eta$ approaches the kinematic limit of $2/3$.  

Since the typical DM velocity dispersion in the Galaxy today is $\< v^2\> =\mathcal{O} (10^{-6} )$, whereas $\< v^2\> =\mathcal{O}( 10^{-1} )$ at $x_f$, the \textit{p}-wave cross-section today is suppressed by five to six orders of magnitude relative to its value at freezeout. Figure~\ref{fig:CrossSection} thus demonstrates that DM annihilations in the galaxy today are dominated by the $s$-wave $\chi\chi\to aaa$ process. 
We observe two distinct regimes, depending on whether the $p$- or $s$-wave process dominates at $x_f$.  When the $p$-wave process dominates, at small DM mass and large $\eta$, the $s$-wave annihilation cross-section is suppressed by orders of magnitude compared to the typical thermal target ($\sim 1$ pb).  On the other hand, when the $s$-wave cross-section dominates, at large DM mass and small $\eta$, it is comparable to the thermal target.

\subsubsection{Pseudoscalar decays into SM final states}
\label{sec:decays}

The pseudoscalar $a$ can decay to the SM through dimension-five axion-like couplings to SM gauge bosons. For simplicity, we will consider here the case when the pseudoscalar couples at leading order only to the hypercharge field strength $B^{\mu\nu}$ via the interaction 
\beq
\label{eq:ldecay}
\frac{a}{\Lambda}  \epsilon_{\mu\nu\rho\sigma} B^{\mu\nu}B^{\rho\sigma}.
\eeq
This choice is also an interesting scenario for discoverability, as it yields an energetic gamma-ray spectrum with a distinctive spectral feature.

After electroweak symmetry breaking, the interaction of Eq.~\ref{eq:ldecay} mediates the decays $a\to \gamma\gamma$ and, if kinematically possible, $a\to Z\gamma$ and $a\to ZZ$.
For a given value of $m_a$, the partial widths into $\gamma\gamma$, $Z\gamma$ and $ZZ$ final states following from the interaction of Eq.~\ref{eq:ldecay} are given by
\begin{eqnarray}
\label{eq:brs}
&\Gamma& (a\to \gamma\gamma) =  \frac{C}{2}\cos^2\theta_W, \\
\nonumber
&\Gamma& (a\to Z\gamma) =  2 C\cos\theta_W\sin\theta_W\left( 1 -\frac{m_Z^2}{m_a^2}\right)^3 \Theta(m_a-m_Z), \\
\nonumber
&\Gamma& (a\to Z Z) =\frac{C}{2}\sin^2\theta_W \left( 1 -\frac{4m_Z^2}{m_a^2}\right)^{3/2}  \Theta(m_a-2m_Z),
\end{eqnarray}
where $\theta_W$ is the Weinberg angle and $C\sim m_a/\Lambda^2$ is a common constant of proportionality. We require $C$ to be small enough that annihilations of DM directly to SM gauge bosons through an intermediate $a$ are negligible in comparison to the secluded annihilation processes of Fig.~\ref{fig:FeynmanDiagram(LaTeX)}, but otherwise our results do not depend on $C$. We require only the relative branching fractions of the pseudoscalar into the various decay channels that follow from Eq.~\ref{eq:brs}. 

Finally, it is worth observing that, unlike traditional WIMP models, the final states in this model are dominated by photons.  In theories that produce copious amounts of charged particles in DM annihilations, secondary synchrotron radiation can provide a competitive probe of DM annihilations within a spike, thanks in large part to the better angular resolution afforded by the lower-energy photons \cite{Bertone:2002je,Aloisio:2004hy, Regis:2008ij}, although the relative magnitude of this synchrotron signal depends on relatively uncertain aspects of the modeling of the Galactic Center \cite{Cholis:2014fja}. However, in the nightmare dark matter model studied here,  DM annihilations proceed directly to gamma rays in the mass range $m_\chi \sim \mathcal{O}(100\gev)$ where radio constraints are especially relevant.  Thus the primary signal of this model is in gamma rays, where the photon spectrum exhibits a prominent and distinctive feature, as we discuss next.

\subsection{Photon Spectra}

In this subsection we determine the photon spectra $dN/dE_\gamma$ resulting from DM annihilations $\chi \chi \rightarrow aaa$. 

\subsubsection{Photon spectra from $a \rightarrow \gamma \gamma$ decays}

We begin with the photon spectrum $dN/dE_\gamma$ from an \textit{s}-wave annihilation process $\chi \chi \rightarrow aaa \rightarrow 6 \gamma$, which can be obtained analytically in the non-relativistic limit. Consider an individual $a$ particle with energy $E_a$ in the Galactic frame decaying into two photons. The maximum and minimum energies the photons can have are 
\begin{gather}
E_{\text{max/min}}=\frac{1}{2}\Big(E_a \pm \sqrt{E_a^2-m_a^2} \Big).
\label{eqn:ERange}
\end{gather}
As the decays of $a$ are isotropic in its rest frame, the energy distribution of daughter photons
is uniform between the kinematic boundaries. Defining the dimensionless variables
\begin{gather}
u \eee \frac{E_\gamma}{m_\chi} = \frac{E_\gamma}{E_a} \frac{E_a}{m_\chi}
\label{eqn:GammaDefinition}
\end{gather}
and  
$w\equiv E_a/m_\chi$, 
the kinematic endpoints can be written
\begin{gather}
u_{\text{max/min}}=\frac{1}{2}\Big(w \pm \sqrt{w^2-\eta^2} \Big).
\label{eqn:GammaRange}
\end{gather}
For a given $w$ and $\eta$, the probability $P(u;w, \eta) du$ of finding a photon within the energy interval $du$ is thus
\begin{gather}
P(u;w,\eta) du = \frac{du}{(u_{\text{max}}-u_{\text{min}})} =  \frac{du}{\sqrt{w^2-\eta^2}},
\label{eqn:Probablity1}
\end{gather}
as the probability is uniform and unit-normalized over the kinematically allowed interval. The probability of obtaining an $a$ particle with (relative) energy $w$ from the process $\chi\chi\to aaa$ is given by
\begin{eqnarray}
P(w;\eta)&=& \frac{1}{\sigma} \frac{d\sigma}{d w} =  A \int_{w_{2-}}^{w_{2+}} \,  dw_2 f(w, w_2;\eta)\\
      &\equiv& A f(w ; \eta),
\end{eqnarray}
where the cross-section $\sigma$ for $\chi\chi\to aaa$ is given in Eq.~\ref{eqn:sCrossSection} and $A$ is a normalization factor.
%
\begin{figure}
	\includegraphics[scale=0.65]{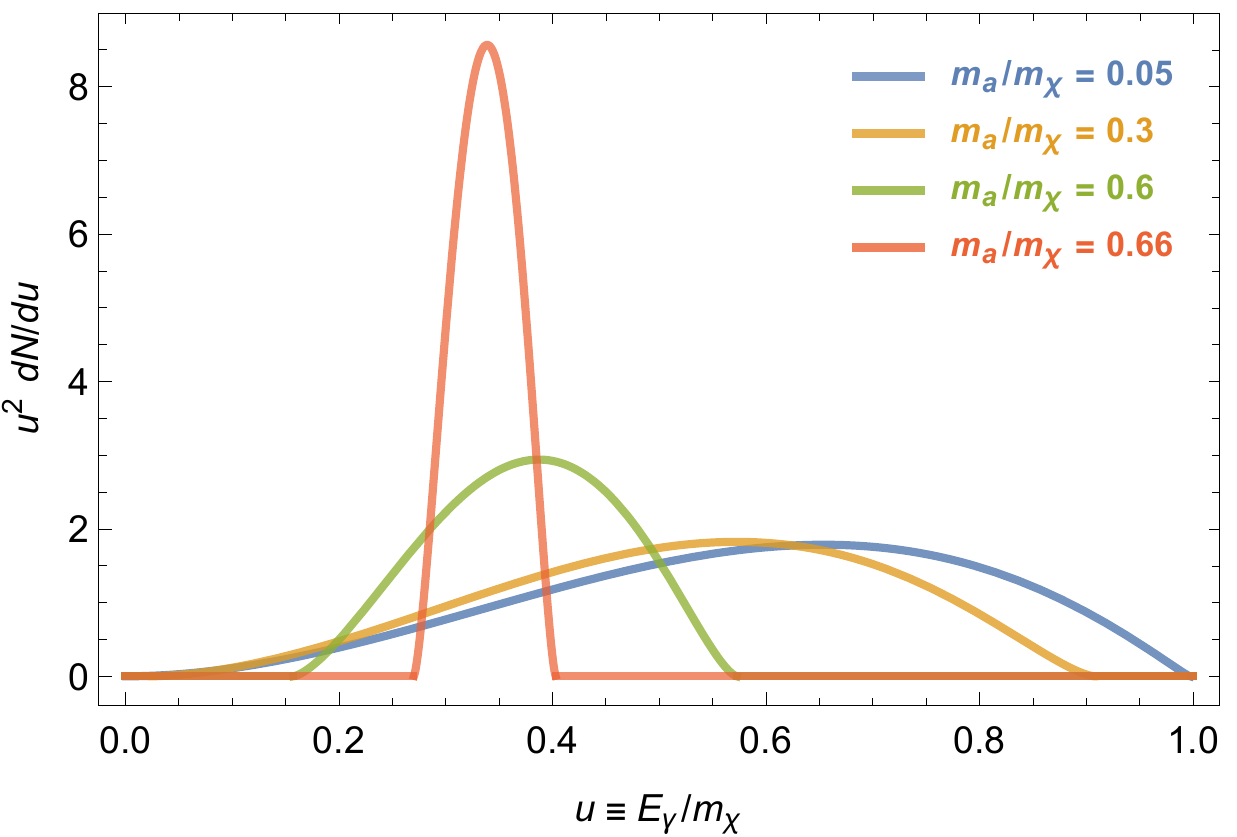}
	\caption{Photon spectra resulting from $\chi\chi\to aaa\to 6\gamma$ in the HSAP model, for four fiducial values of $\eta=m_a/m_\chi$: 0.05 (blue), $0.3$ (yellow), $0.6$ (green), and $0.66$ (red). As $\eta$ increases toward the kinematic limit, the restricted phase space forces the spectrum to become increasingly narrow and peaked.}
	\label{fig:EnergyWeightedPhotonEnergy}
\end{figure}
The probability density of finding any specific combination of $u, w$ is then
\begin{gather}
P( u, w; \eta) = A \frac{f(w; \eta)}{\sqrt{w^2-\eta^2}}.
\label{eqn:Probablity2}
\end{gather}
As this expression indicates, a daughter photon with energy $u$
may have come from a parent $a$ with a range of possible energies $w$. To obtain the probability of observing a photon with energy $u$, we integrate $P(u, w;\eta)$ over the range of $w$ consistent with the value of $u$. The maximum possible value of $w$ is, from Eq.~\ref{eqn:w1Limits}, $w_\text{max} = 1 - \frac{3}{4}\eta^2$,
independent of $u$. For a fixed value of $u$, $w_\text{min}$ can be determined from Eq.~\ref{eqn:GammaRange}, which gives
\begin{gather}
w_\text{min}(u) = \frac{4u^2+\eta^2}{4u}.
\label{eqn:xLowerBound}
\end{gather}
The desired photon spectrum is therefore
\begin{gather}
\frac{dN}{dE_{\gamma}}(u,\eta) = 6 A\int_{w_\text{min}}^{w_\text{max}} \frac{w_\text{max}(\eta)}{w_\text{min}(u,\eta)} \frac{f(w;\eta)}{\sqrt{w^2-\eta^2}}  dw.
\label{eqn:dNdE1}
\end{gather}
The factor of six appears here since there are six final state photons in the annihilation $\chi \chi \rightarrow aaa \rightarrow 6 \gamma$. Accordingly, we evaluate the normalization factor $A$ by requiring 
\begin{gather}
\int_{u_\text{min}}^{u_\text{max}} \frac{dN}{dE_{\gamma}}(u,\eta) dE_{\gamma} = 6.
\label{eqn:dNdE2}
\end{gather}

In Fig.~\ref{fig:EnergyWeightedPhotonEnergy}, the normalized photon spectra, weighted by $u^2$, are plotted for four different values of $\eta$. For small $\eta$, the energy distribution is broad and peaks at relatively high energies compared to the spectra for larger values of $\eta$. As $\eta$ approaches the kinematic limit of $2/3$, the changes in the spectrum shape become increasingly rapid as the available phase space shrinks.

\subsubsection{Photon spectra from $a \rightarrow Z\gamma$ and $a \rightarrow ZZ$ decays}
%
\begin{figure}
	\includegraphics[scale=0.6298]{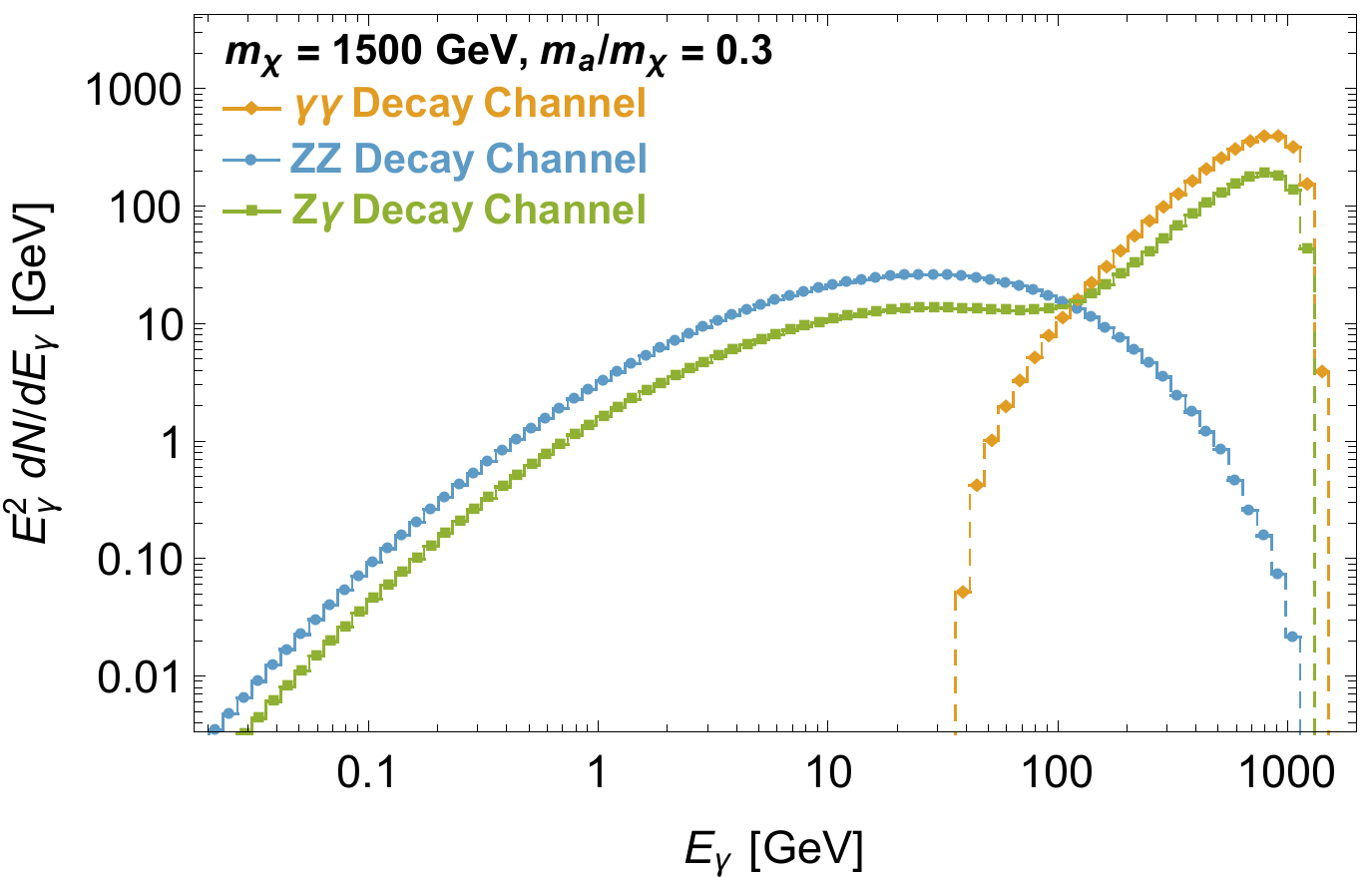}
	\caption{Photon spectra $E^2 dN/dE$ resulting from $a$ decays to $\gamma \gamma$ (yellow), $ZZ$ (blue), and $Z\gamma$ (green). We fix $m_\chi = 1500$ GeV and $\eta = 0.3$.
}
	\label{fig:DecayChannelEnergyDist}
\end{figure}
Once $m_a > m_Z$ ($2m_Z$), the decay channel $a\to Z\gamma$ ($a\to ZZ$) opens up. These decay channels result in a continuum of numerous but lower-energy photons from hadron decays, as well as final state radiation off of charged leptons. To obtain photon spectra $dN/dE$ for DM annihilation channels with any number of final-state $Z$ bosons, we proceed numerically. A Monte Carlo sampling method was employed to compute the four-momenta of $a$ particles produced in $\chi\chi\to aaa$ annihilations following the non-relativistic distribution of Eq.~\ref{eqn:PhaseSpace}, as well as the momenta of their daughter photons and $Z$ bosons. 
 The photon spectrum resulting from a $Z$ boson in its rest frame was computed using \texttt{Pythia 8} \cite{Sjostrand:2007gs}. The resulting photons were then boosted to the Galactic rest frame.

Figure~\ref{fig:DecayChannelEnergyDist} shows the normalized photon spectra resulting from $a$ decays to $\gamma \gamma$ (yellow), $ZZ$ (blue), and $Z\gamma$ (green) for an example parameter point with $m_\chi = 1500$~GeV and $\eta = 0.3$. The $\gamma \gamma$ decay process produces comparatively more high-energy photons with a narrower distribution. The spectrum from $ZZ$ decay is considerably broader with a smooth low-energy tail. The $Z\gamma$ decay spectrum inherits features from both $\gamma \gamma$ and $ZZ$ spectra, albeit with a slightly smaller maximum photon energy than the $\gamma \gamma$ spectrum due to the non-zero $m_Z$. 

\section{The Astrophysical Model}
\label{sec:spike}

Following \cite{Fields:2014pia, Shelton:2015aqa, Johnson:2019hsm}, we consider a simple parametric model describing possible DM density spikes at the Galactic Center.
We take the DM halo to be described by a generalized Navarro-Frenk-White (NFW) model, which in the central regions of the Galaxy, i.e., well within the scale radius, can be described by a power law $\rho(r) = \rho(r_0) (r/r_0)^{\gamma_\text{c}}$. We anchor this power law to the Solar System, where we take the local density of
DM to be $\rho(r_\odot) = 0.3\,\mathrm{GeV}/{\mathrm{cm}^3}$
\cite{Bovy:2012tw}, at a distance $r_{\odot}=8.46$~kpc
from the Galactic Center \cite{Do:2013upa}. DM-only simulations typically yield values of the cusp exponent $\gamma_\text{c}$ within the range $0.9\lesssim \gamma_\text{c}\lesssim 1.2$ \cite{Diemand:2008in,Navarro:2008kc}, while the adiabatic contraction of the central halo following baryonic collapse can lead to larger values of $\gamma_\text{c}$ \cite{Blumenthal:1985qy,Gnedin:2004cx,Gustafsson:2006gr}. 
We will treat $\gamma_\text{c}$ as a free parameter.

The Milky Way's SMBH has mass $M= 4\times 10^6 M_\odot$ \cite{Genzel:2010zy,Ghez:2008ms}. Its gravitational zone of influence extends out to the radius
$r_h \equiv G M/v_0^2$, where the gravitational potential energy due to the BH is equal to the typical kinetic energy of a DM particle in the halo. Here $v_0$ is the velocity dispersion in the inner halo. We adopt as our fiducial dispersion $v_0 = 105 \pm 20 {\rm ~km~s^{-1}}$~\cite{Gultekin:2009qn}. Numerical studies indicate that the spike itself begins growing somewhat within $r_h$, at $r_b = 0.2 r_h$ \cite{Merritt:2003qc,Merritt:2003qk}. The spike is also well-described as a power law, $\rho_\text{sp} (r) = \rho(r_b) (r/r_b)^{\gamma_\text{sp}}$, although the spike index $\gamma_\text{sp}$ depends sensitively on the formation history of the SMBH and the properties of its environment.
Spikes that form around a BH that grows adiabatically at the center of a cuspy DM halo are very steep, $\gamma_{\text{sp}} = (9 - 2 \gamma_{\text{c}})/(4 - \gamma_{\text{c}})$ \cite{Gondolo:1999ef}; on the other hand, if the BH is not at the dynamical center of its halo, then it produces a very shallow spike, $\gamma_\text{sp} = 1/2$ \cite{Nakano:1999ta,Ullio:2001fb}. The dynamical heating of DM from gravitational interactions with a dense and cuspy stellar distribution results in a spike solution with limiting index $\gamma_\text{sp} = 1.5$, attained when the system has fully equilibrated \cite{Merritt:2003qk, Gnedin:2003rj, Bertone:2005hw, Merritt:2006mt}. Non-equilibrated spikes with intermediate power laws are possible if the DM at the Galactic Center is still in the process of equilibrating \cite{Merritt:2003qk}.\footnote{An alternate parameterization of non-equilibrated spikes undergoing a baryonic heating process, developed in \cite{Ahn:2007ty}, models the reduction of the spike signal through modifications of $r_b$ rather than $\gamma_\text{sp}$.  Both approaches give indistinguishable results for the spike signal \cite{Sandick:2016zeg}.} In order to succinctly capture the range of signal strengths predicted by these various different spike formation and evolution scenarios, we treat $\gamma_\text{sp}$ as a free parameter and vary it between $1.5$ and its adiabatic value.

Once the DM density in the spike reaches the value $\rho_\text{ann} = m_\chi/(\langle\sigma v\rangle \tau)$, where $\tau \approx 10^{10}$ yr is the lifetime of the spike, DM annihilations become rapid enough to deplete the spike. We define the radius at which this occurs as $r_\text{in} = r_b \cdot (\rho_b/\rho_\text{ann})^{1/\gamma_{\text{sp}}} $. For $r<r_\text{in}$, the spike follows a very mild power law, $\rho(r)\propto r^{-1/2}$ in the case of $s$-wave annihilations \cite{Vasiliev:2007vh,Shapiro:2016ypb}. Finally the inner boundary of the spike is obtained at $r < 4 G M$ \cite{Sadeghian:2013laa}.
Altogether, then, we take for our model
\begin{eqnarray}
\rho(r) &=&0 \;\;\;\;\;\;\;\;\;\;\;\;\;\;\;\;\;\;\;\;\;\;\;\;\;\;\;\;\;\;\;(r < 4GM), \\\nonumber
& =& \frac{\rho_{\text{sp}}(r)\rho_{\text{in}}(r)}{\rho_{\text{sp}}(r)+\rho_{\text{in}}(r)} \;\;\;\;\;\;\;\;\;(4GM \leq r < r_{b}), \\\nonumber
&=&\rho_{b}(r_{b}/r)^{\gamma_{\text{c}}}\;\;\;\;\;\;\;\;\;\;\;\;\;\;\;\;\;(r_{b} \leq r < r_\odot) ,
\label{eqn:DensityProfile}
\end{eqnarray}
where $\rho_{\text{sp}}(r) = \rho_{b}(r_{b}/r)^{\gamma_{\text{sp}}}$, $\rho_{\text{in}}(r) = \rho_{\text{ann}}(r_{\text{in}}/r)^{0.5}$, $ \rho_{b} = \rho(r_{\odot}) \cdot (r_\odot/r_{b})^{\gamma_{\text{c}}}$, and 
we have defined a spike profile that smoothly interpolates between the inner spike with index $1/2$ and the outer spike with index $\gamma_\text{sp}$.

With the parameter values adopted here, $r_b~\approx 0.3$~pc, while for typical DM parameters in the HSAP model the radius $r_\text{in}\sim \mathrm{few}\times10^{-5}$~pc.
The inner spike structure, which dominates the emission, then subtends $\sim 4$ milliarcseconds on the sky, which is several orders of magnitude below the typical resolution of the \textit{Fermi} telescope. We thus treat the spike signal as a point source.
The differential photon flux observed on Earth from the spike is then given by
\begin{gather}
\frac{d\Phi_\text{sp}}{dE_{\gamma}} = \frac{1}{r_\odot^2}\int_{4GM}^{r_\text{b}} dr \rho(r)^2 r^2 \frac{\< \sigma v \>}{2m_{\chi}^2}\frac{dN}{dE_{\gamma}}.
\label{eqn:SpikeFluxFunction}
\end{gather}
However, the spike sits on top of a bright halo, which can also contribute to the observed central signal as
\begin{gather}
\frac{d\Phi}{dE_{\gamma}} = \int_{\Delta \Omega}d\Omega \int_{\text{l.o.s.}} \rho(r)^2 \frac{1}{4\pi}\frac{\< \sigma v \>}{2m_{\chi}^2}\frac{dN}{dE_{\gamma}} d\ell,
\label{eqn:FluxFunction}
\end{gather}
where the line-of-sight (l.o.s.) distance $\ell$ at an angle $\theta$ away from the GC is given by
\begin{gather}
r = \sqrt{\ell^2 + r_\odot^2 - 2 \ell r_\odot \cos\theta}.
\label{eqn:losDistance}
\end{gather}
We conservatively include the emission from the central region of the halo, using an energy-dependent half-angle for  \textit{Fermi} and $0.5$\degree\ for H.E.S.S., and indicate where this halo emission, $\Phi_{\mathrm{halo}},$ exceeds the contribution from the spike, $\Phi_{\mathrm{sp}}$ in the dominant energy bin.  Our main interest is in the regime where the spike dominates, $\Phi_{\mathrm{sp}}\gg \Phi_{\mathrm{halo}}$. This choice allows us to succinctly account for both the finite angular resolution of gamma-ray telescopes, notably \it{Fermi} \cite{fermipsf}, and demarcate the region where we might expect challenges from resolving a point source on top of a bright and non-uniform background. 
 To account for \it{Fermi}'s energy-dependent angular resolution, we consider the halo flux within a cone with half-angle given by the average 68\% containment angle as a function of photon energy, which ranges between $5.3\degree$ for 100 MeV photons and $0.1\degree$ for 100 GeV photons \cite{fermipsf}.  
In plotting spike spectra and evaluating limits, in each bin we adopt an effective angular resolution using the 68\% containment angle of the average photon energy in that bin, as determined from the $\gamma\gamma$ energy spectrum, Eq. \ref{eqn:dNdE1}.

\section{Sensitivity to faint $s$-wave annihilation signals}
\label{sec:data}

\begin{figure}
\includegraphics[width=\linewidth]{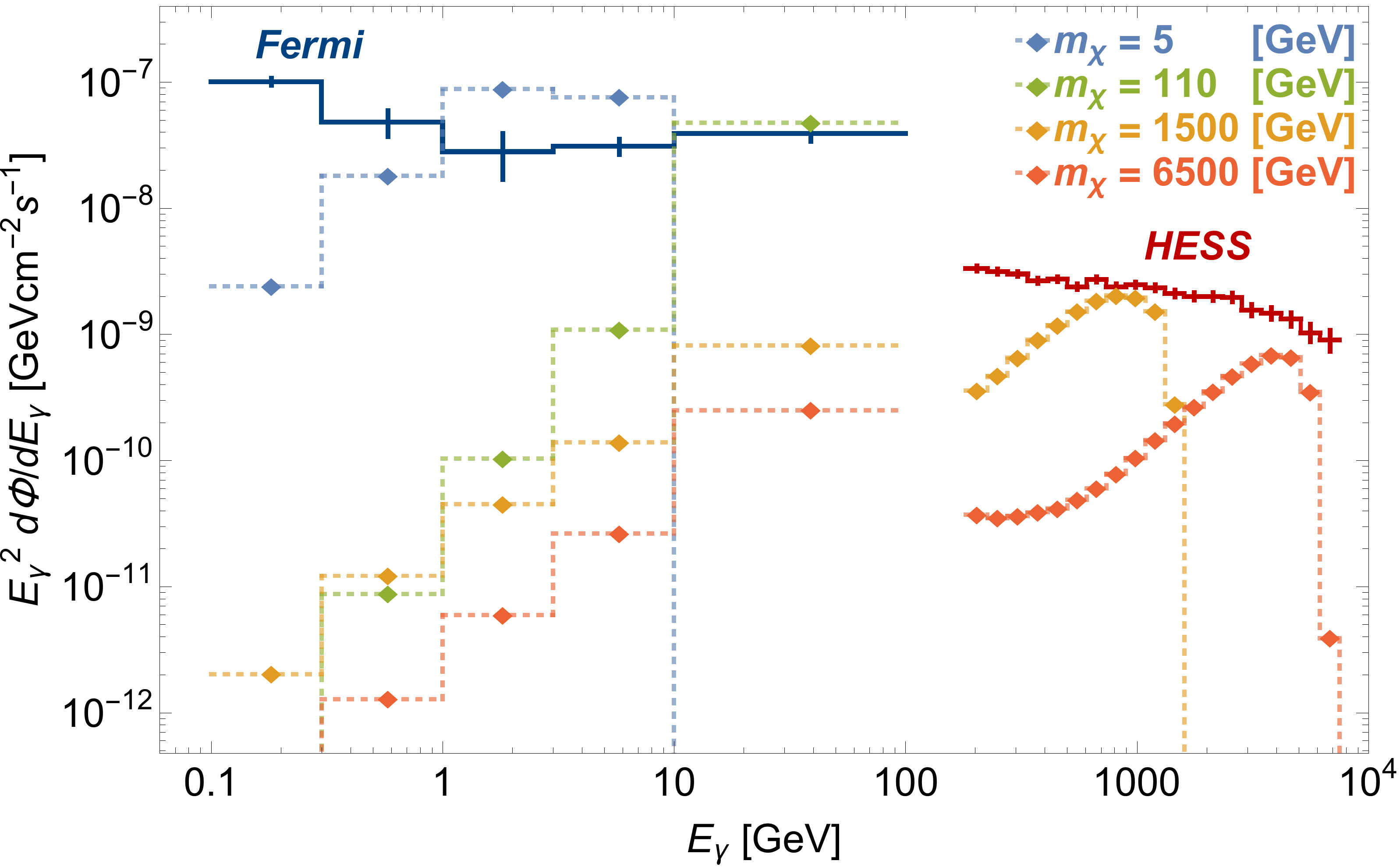}
\caption{Observed GC $\gamma$-ray point source spectra compared with example model predictions. The spectra of \textit{Fermi} source \texttt{3FGL J1745.6-2859c} (dark blue) \cite{Acero:2015hja} and H.E.S.S. source \texttt{HESS J1745-290} (dark red) \cite{Abramowski:2016mir} associated with Sgr A* are plotted in solid. The dashed histograms are four predicted flux spectra for $m_\chi = 5$ GeV (light blue), $110$ (green), $1500$ (yellow), and $6500$ GeV (orange). We fix $\gamma_\text{c} = 1.2$, $\gamma_\text{sp} = 1.8$, and $m_a/m_\chi = 0.1$. 
}
\label{fig:FermiHESS}
\end{figure}

Both \textit{Fermi} and H.E.S.S. have observed bright point sources in the Galactic Center that the respective collaborations have associated with Sgr A*. The \textit{Fermi} point source \texttt{3FGL J1745.6-2859c} is observed in gamma rays with $100 \,\mathrm{MeV} < E_\gamma<100$ GeV \cite{Acero:2015hja}, while the H.E.S.S. source \texttt{HESS J1745-290} is observed in gamma rays with $180 \,\mathrm{GeV} < E_\gamma< 79$ TeV  \cite{Abramowski:2016mir}.  We show the observed spectra of these point sources in the solid histograms of Figure~\ref{fig:FermiHESS}. 
The abrupt decrease in binned flux magnitude starting at $m_\chi = 180$ GeV reflects the different bin sizes in the two experiments. The energy gap between \textit{Fermi} and H.E.S.S. datasets at $100 < m_\chi < 180$~GeV is clearly visible.
For comparison, the dashed histograms show four example predictions for the primary photon spectra arising from HSAP DM annihilations within DM density spikes.  We show predictions for four different values of DM mass at fixed $\eta = 0.1$, $\gamma_\text{c} = 1.2$, and $\gamma_\text{sp} = 1.8$. 
%

\begin{figure}
		\includegraphics[width=\linewidth]{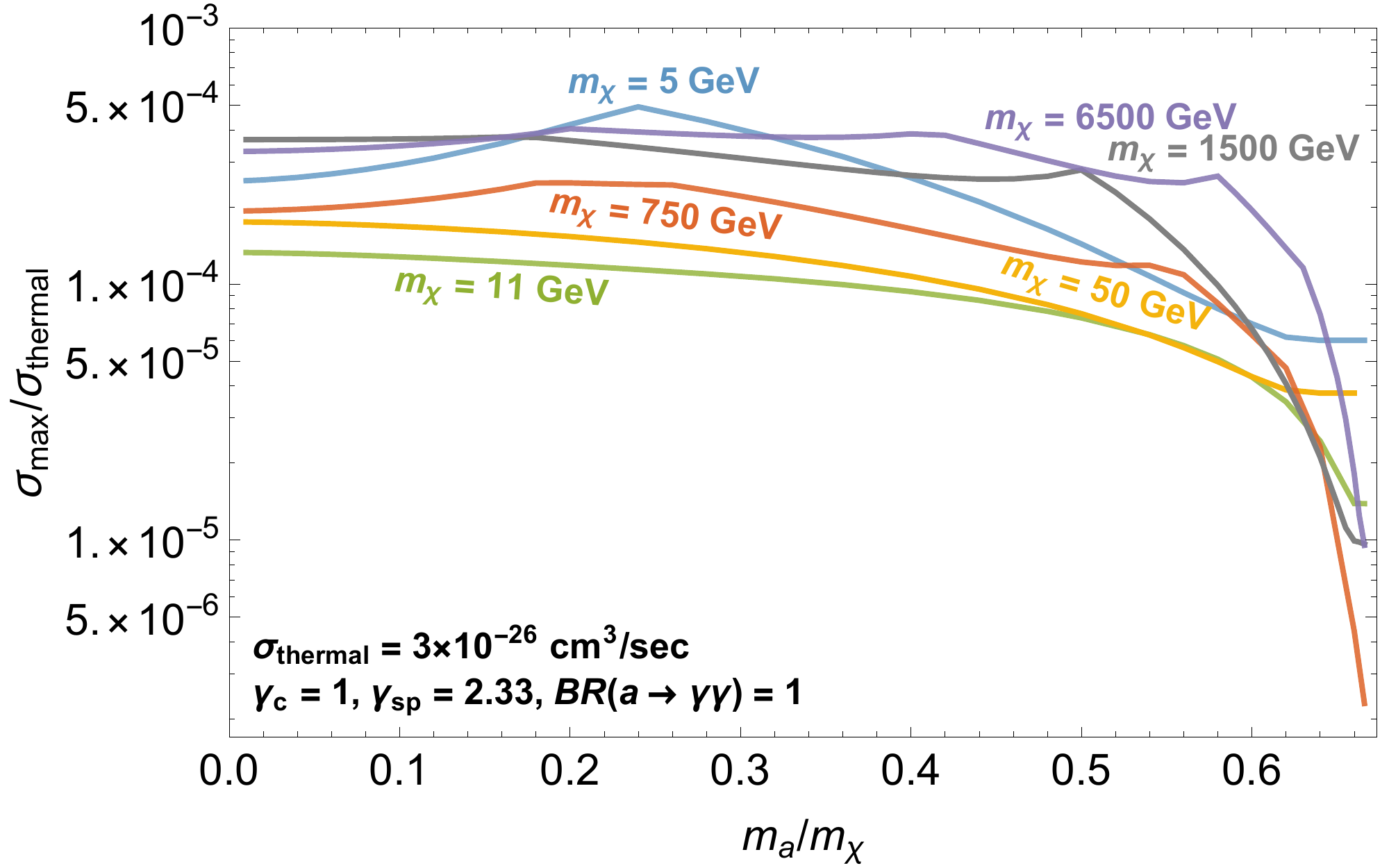}
	\includegraphics[width=\linewidth]{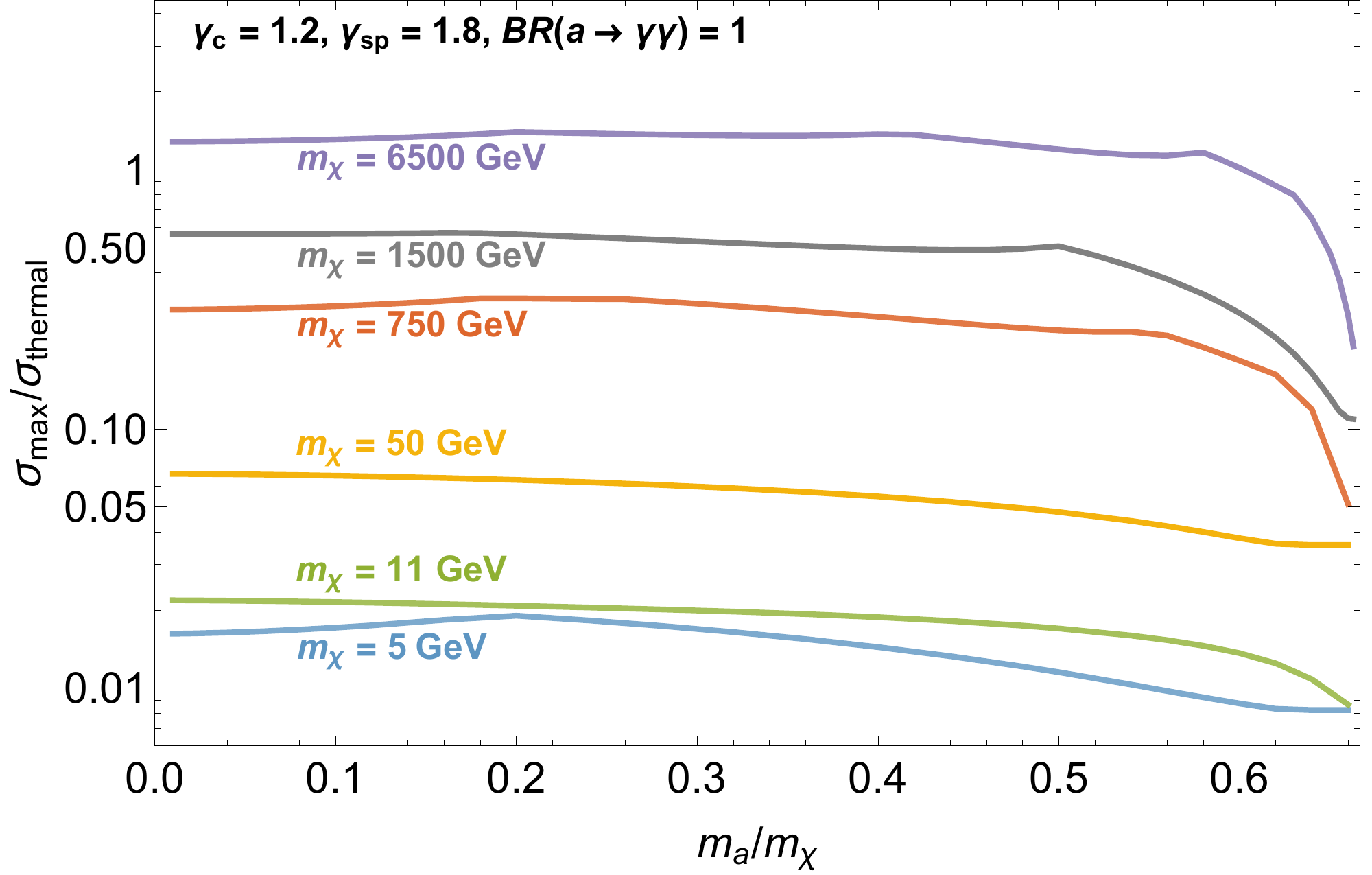}
	\caption{Value of the $s$-wave cross-section, as a fraction of the reference thermal value $\sigma_{\mathrm{thermal}} \equiv 3\times 10^{-26}\mathrm{cm}^3/$s, for which the photon flux from a BH spike exceeds either \textit{Fermi} or H.E.S.S. observations at 95\% CL in at least one bin. Results are shown as a function of the mass ratio $\eta =m_a/m_\chi$ for six fixed values $m_\chi$. In the top panel we fix $\gamma_\text{c} = 1$ and consider the adiabatic $\gamma_\text{sp} = 2.33$, and in the bottom panel we use $\gamma_\text{c} = 1.2 $ and $\gamma_\text{sp} = 1.8$.  We take $Br(a\to\gamma\gamma)=1$.
}
	\label{fig:XsecsExclusion}
\end{figure}

%
To estimate sensitivity to HSAP DM annihilations within BH spikes, we use the simple criterion that the spike flux should not exceed the observed flux from either \texttt{3FGL J1745.6-2859c} or \texttt{HESS J1745-290} at more than 95\% CL in any bin.  Figure~\ref{fig:XsecsExclusion} shows the values of the $s$-wave annihilation cross-section excluded by this procedure for six fixed values of $m_\chi$ in two different astrophysical scenarios. In this figure we take $Br(a\to\gamma\gamma)=1$ throughout, so that the photon spectrum is given by Eq.~\ref{eqn:dNdE1}.  Visible kinks in the lines reflect when the peak of the photon spectrum moves from one bin to another as $\eta$ changes.  For $m_\chi=5,$ 11, and $50$ GeV, the constraints come from \textit{Fermi}, while for $m_\chi =750,$ 1500, and 6500~GeV, the constraints come from H.E.S.S.  The improvement of the limits as $\eta\to 2/3$ reflects the increasing sharpness of the peak in the photon spectrum. From Fig.~\ref{fig:XsecsExclusion}, we see that adiabatic spikes are sensitive to cross-sections some four orders of magnitude smaller than the standard thermal target, while for more moderate values of $\gamma_\text{sp}$, the resulting sensitivity even in relatively cuspy haloes is less dramatic but still interesting.  
Constraints on the cross-section for higher mass DM are generally weaker; 
however, in the HSAP model studied in this paper, this effect is substantially offset by the larger predicted $s$-wave cross-sections at high mass.

\begin{figure}
\includegraphics[width=\linewidth]{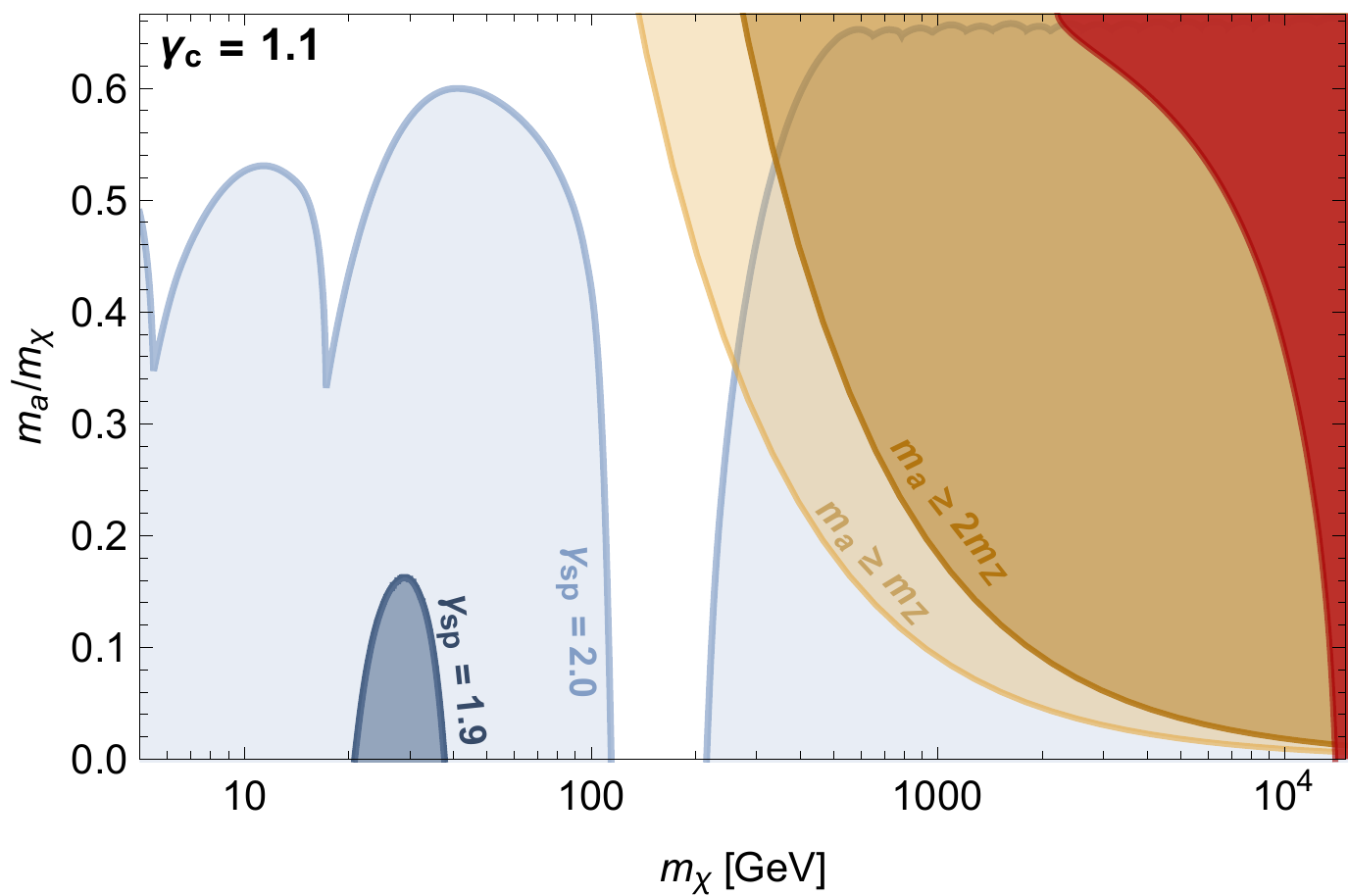}
\includegraphics[width=\linewidth]{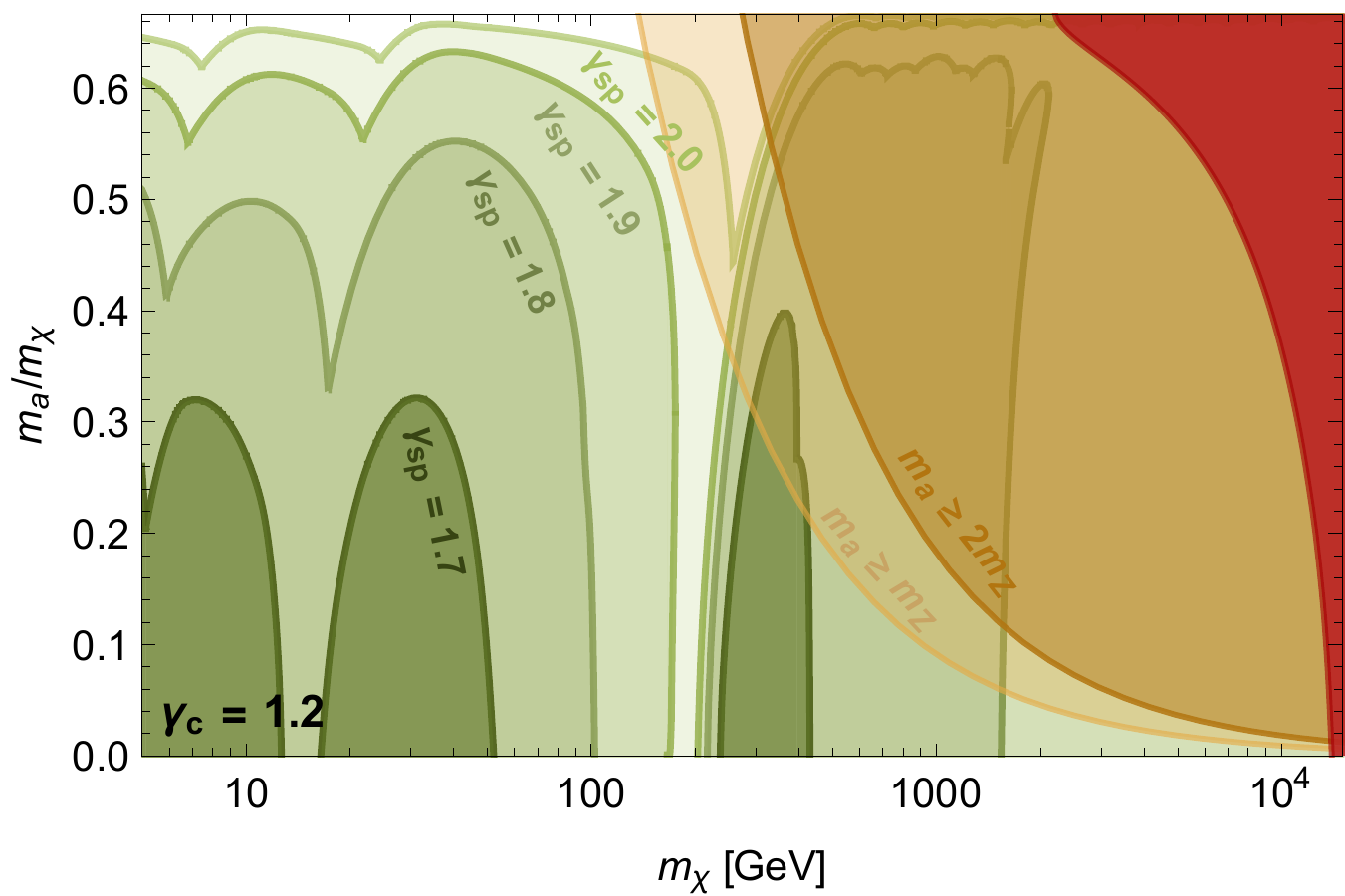}
	\caption{Excluded regions in $m_a$\textendash$m_\chi$ parameter space for fixed $\gamma_\text{c}$, $\gamma_\text{sp}$. The top panel fixes $\gamma_{\text{c}} = 1.1$ and $\gamma_{\text{sp}} = 1.9$, 2.0; the bottom takes $\gamma_{\text{c}} = 1.2$ and $\gamma_{\text{sp}} = 1.7$, 1.8, 1.9, 2.0. We take $Br(a \rightarrow \gamma \gamma)=1$ throughout.  The $a \rightarrow Z \gamma$ ($a\to Z Z$) decay channel is available in the light (dark) tan shaded regions.  The dark red region  
indicates where $y^2/4\pi \geq 1$.
}
	\label{fig:MchiMaExclusion}
\end{figure}

Figure~\ref{fig:MchiMaExclusion} shows excluded regions of $m_\chi$\textendash$\eta$ parameter space for fixed choices of spike and halo indices. 
The top panel fixes $\gamma_{\text{c}} = 1.1$ together with $\gamma_{\text{sp}} = 1.9$, $2.0$ (no parameter space is excluded for $\gamma_{\text{sp}} \leq 1.8$). The bottom panel has $\gamma_{\text{c}} = 1.2$ with $\gamma_{\text{sp}} = 1.7$, $1.8$, $1.9$, $2.0$. For each exclusion boundary, a concave curve segment corresponds to the exclusion given by one energy bin from either the \textit{Fermi} or H.E.S.S. point source. As expected from Fig.~\ref{fig:CrossSection}, we see strong variation of the sensitivity with the mass ratio $\eta$ at small $m_\chi$.

In Fig.~\ref{fig:MchiMaExclusion} we continue to
take for simplicity \mbox{$Br(a\rightarrow\gamma\gamma)=1$}, as this case can be handled analytically.  We indicate with the shaded light (dark) tan area where $m_a>m_Z\, (2m_Z)$, and this assumption breaks down.   In general, once $Z$ bosons are kinematically accessible, the details of the DM annihilation spectrum will depend on the modeling choice of how the pseudoscalar $a$ couples to SM electroweak bosons.  However it is straightforward to translate the results of Fig.~\ref{fig:MchiMaExclusion} to the non-zero branching fractions into $Z\gamma$ and $ZZ$ predicted by Eq.~\ref{eq:brs} in most of parameter space, as we now discuss.

\begin{figure}
\includegraphics[width=\linewidth]{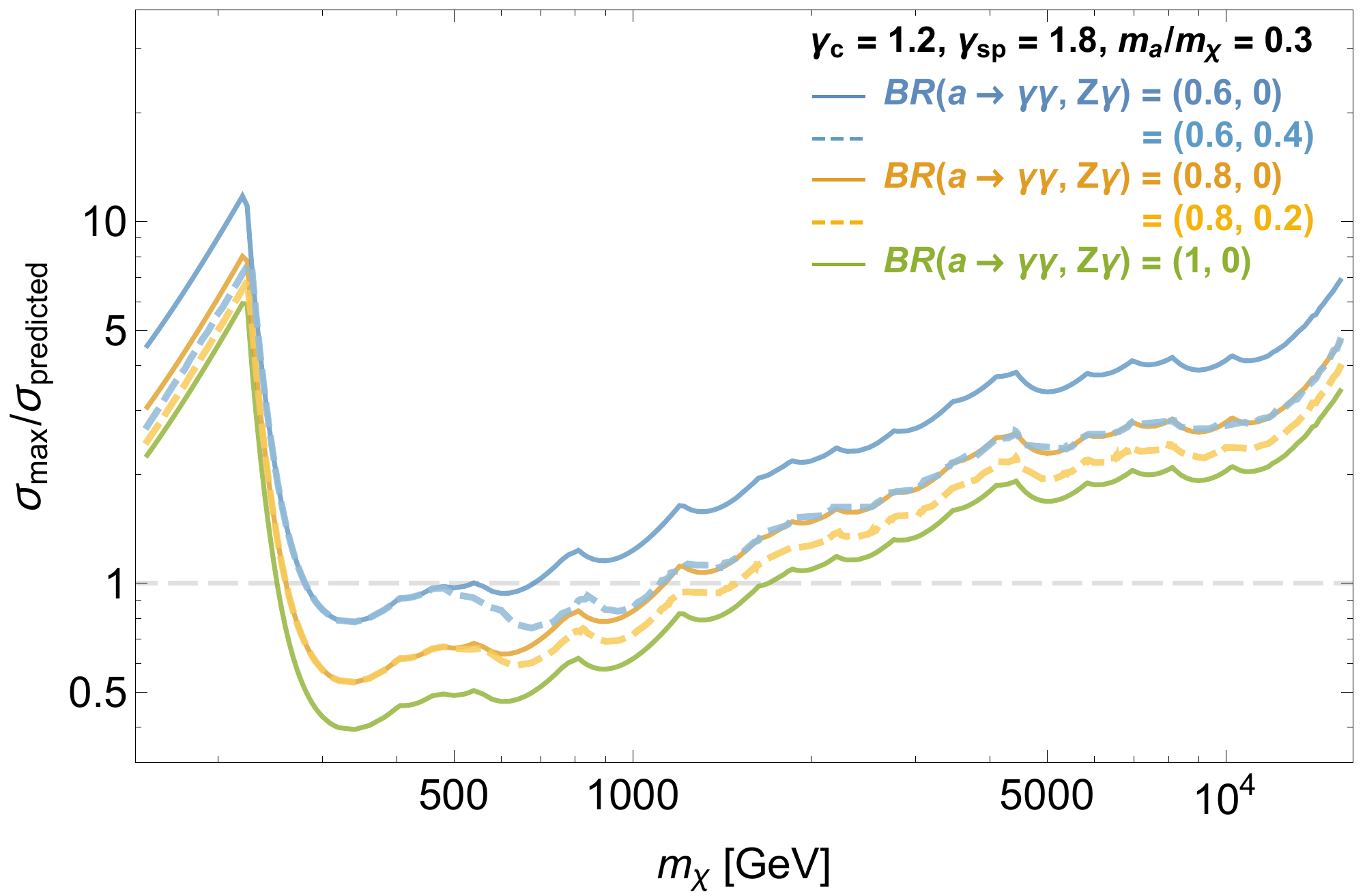}
\caption{Dependence of the maximum allowed $s$-wave annihilation cross-section on the DM annihilation channels.  
We study limits for fixed $\{Br(a \rightarrow \gamma \gamma), Br(a\to Z\gamma)\}$ = $\{1,0\} $, $\{0.8,0.2\}$, and $\{0.6,0.4\}$ in blue, yellow, and green respectively.
Solid lines use only photons from the $\gamma\gamma$ channel, while dashed lines use all photons. Results are shown for an example point with $\gamma_\text{c} = 1.2$, $\gamma_\text{sp} = 1.8$, and $\eta = 0.3$.}
\label{fig:CrossSectionScaling}
\end{figure}

Sensitivity to DM annihilations in the HSAP model is dominated by the peak of the photon spectrum in the $\gamma\gamma$ and $Z\gamma$ channels; the continuum photons coming from $Z$ bosons are not important for determining the sensitivity.   This can be seen from Fig.~\ref{fig:CrossSectionScaling}, where we compare limits set using only $\gamma\gamma$ decays to those determined from both $\gamma\gamma$ and $Z\gamma$ decays.
This figure shows the maximum \textit{s}-wave annihilation cross-section allowed by observations of the GC, relative to the prediction of the HSAP model, as a function of DM mass, and demonstrates how the limit changes as the branching fractions of $a$ are varied.  Green, yellow, and blue curves show exclusions assuming branching ratios $\{Br(a \rightarrow \gamma \gamma), Br(a\to Z\gamma)\}$ = $\{1,0\} $, $\{0.8,0.2\}$, and $\{0.6,0.4\}$ respectively.  The solid curves use only photons from the $\gamma\gamma$ channel to set a limit, while the dashed curves show results from including all photons.\footnote{This discussion is sensible for the SM couplings adopted in this work, where the maximum value of $Br(a\to ZZ)$ is $\approx 12\%$.  In models where $Br(a \to \gamma\gamma)$ is sufficiently suppressed compared to continuum decay modes that the continuum, rather than the peak, dominates the limits, this argument will no longer apply.}

Figure~\ref{fig:CrossSectionScaling} restricts attention to the range of DM masses where $Z$ bosons are kinematically accessible for the chosen value of $\eta =0.3$.  The sharp feature near \mbox{$m_\chi\sim 200$~GeV} reflects the transition from \textit{Fermi} to H.E.S.S.  In the high-$m_\chi$ region, where the peak of the photon spectrum is fully within H.E.S.S.' energy range, the difference in the peak energies of the $\gamma\gamma$ and $Z\gamma$ spectra is small compared to the peak energy.   Thus in the high-$m_\chi$ regime a limit can be estimated simply from counting the number of primary photons from $a$ decays: i.e., by using the $\gamma\gamma$ limit but multiplying by the factor $Br(\gamma\gamma)+\frac{1}{2} Br(Z\gamma) <1$.  This is demonstrated in Fig.~\ref{fig:CrossSectionScaling} by the excellent agreement of the blue dashed and orange solid curves at large $m_\chi$.  Figure~\ref{fig:CrossSectionScaling} also shows the expected {\em linear} scaling of the high-$m_\chi$ excluded cross section with $Br(\gamma\gamma)$.   By contrast, in the gap between \textit{Fermi} and H.E.S.S.' observations, the limit is dominated by the high-energy shoulder of the photon spectrum, where photons from $Z\gamma$ decays make a negligible contribution.

\begin{figure}
\includegraphics[width=\linewidth]{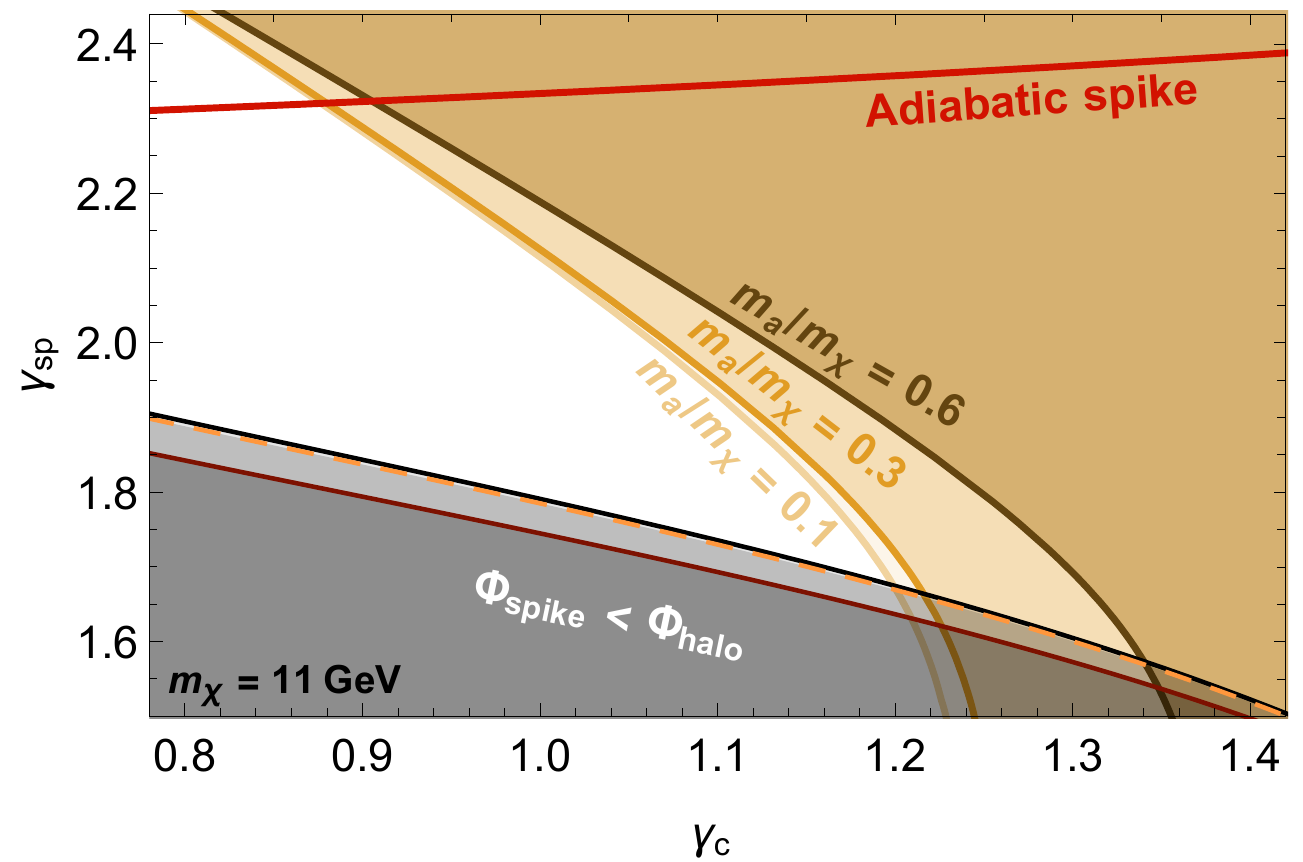}
\includegraphics[width=\linewidth]{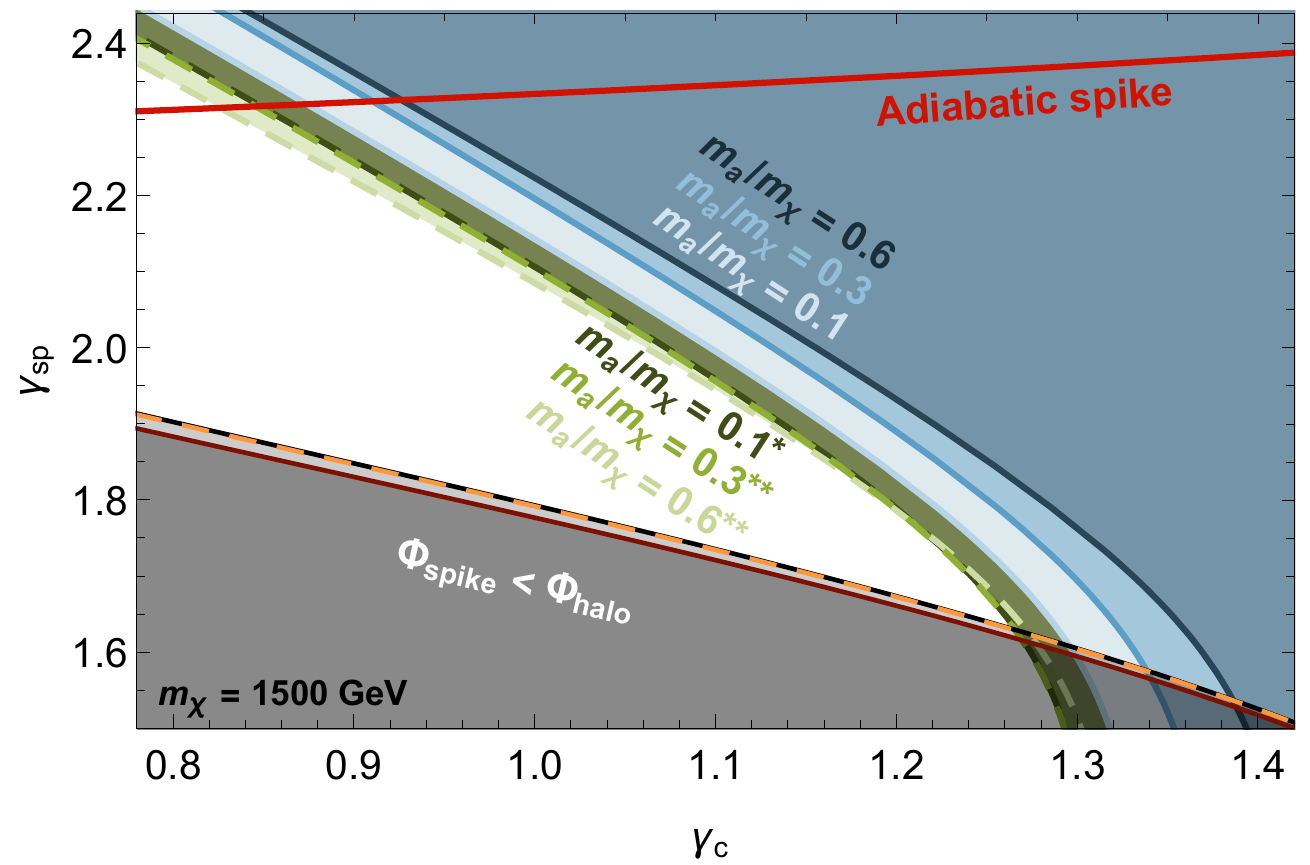}
	\caption{Exclusions in $\gamma_\text{c}$\textendash$\gamma_\text{sp}$ parameter space for fixed $m_\chi =11$ GeV (top) and $1500$ GeV (bottom) and $m_a/m_\chi = 0.1, 0.3, 0.6$.  
Predictions for adiabatic spikes are indicated by the red line. 
The gray shaded areas indicate where the flux from the central region of the halo exceeds the flux from the spike in the energy bin that dominates the limits; the three boundary curves show $\Phi_\text{spike} = \Phi_\text{halo}$ with $m_a/m_\chi = 0.1, 0.3,$ and $0.6$ from top to bottom (black, dashed orange, dark red). In the bottom panel, the blue curves use the branching ratios of Eq.~\ref{eq:brs}, while the green curves, marked with asterisks, artificially assume $Br(a \rightarrow \gamma \gamma) = 1$.}
	\label{fig:GammacGammaspExclusion}
\end{figure}

As Fig.~\ref{fig:MchiMaExclusion} already demonstrates, observational sensitivity to the HSAP model is strongly dependent on $\gamma_\text{sp}$ and especially $\gamma_{\text{c}}$. Any astrophysical parameter combination with $\gamma_{\text{c}} \geq 1.2$ and $\gamma_{\text{sp}} \geq 2.0$ is almost entirely ruled out. Conversely, very little parameter space is excluded for $\gamma_{\text{c}} \leq 1.1$ and $\gamma_{\text{sp}} \leq 1.8$.  We now turn to examining the sensitivity as a function of $\gamma_\text{c}$ and $\gamma_\text{sp}$ for fixed particle parameters in Fig.~\ref{fig:GammacGammaspExclusion}.  Here we show results for fixed $m_\chi = 11$ GeV~(top) and $1500$~GeV (bottom), and three fixed choices of mass ratio $\eta$. The gray shaded regions indicate where the  contribution from the  central $0.5\degree$ of the halo exceeds the contribution from the spike. The exclusions we find are thus indeed driven by the spike for the bulk of parameter space, except at large values of $\gamma_\text{c}$ where a more careful combined study of the halo plus potential spike point source would be warranted.  
 Canonical adiabatic spikes
are almost entirely disfavored, except at very small values of $\gamma_\text{c}$.  

In general, points with $\eta = 0.1$ see more stringent limits than those with $\eta = 0.3$ and $0.6$.  This occurs for two main reasons. First, in the HSAP model, the magnitude of the $s$-wave  cross-section depends sensitively on $\eta$ (see Fig.~\ref{fig:CrossSection}), especially for the relatively low value $m_\chi = 11$ GeV in the top panel.  Second, once  it is kinematically possible to produce $Z$ bosons, the flux in the photon peak is reduced.  The branching fraction $\gamma\gamma+\frac{1}{2} Z\gamma$ is typically larger for smaller values of $\eta$.  This second effect is relevant in the bottom panel of Fig.~\ref{fig:GammacGammaspExclusion}, where  the pseudoscalar is heavy enough to decay into $Z \gamma$ ($\eta = 0.1$, 0.3) and $ZZ$ ($\eta = 0.6$) final states.  
In this panel we show two sets of limits. The green curves, marked with asterisks, show limits obtained by artificially setting $Br(\gamma \gamma) =1$, while the blue curves show limits using the branching fractions of Eq.~\ref{eq:brs}.   In particular, we set 
$\{Br(a \rightarrow \gamma \gamma), Br(a\to Z\gamma)\} \approx \{0.65, 0.35\} $ for $\eta = 0.1$,
$\{Br(a \rightarrow \gamma \gamma), Br(a\to Z\gamma), Br(a\to ZZ)\} \approx \{0.32, 0.61, 0.07\}$ for $\eta = 0.3$, and $\{Br(a \rightarrow \gamma \gamma), Br(a\to Z\gamma), Br(a\to ZZ)\} \approx \{0.3, 0.62, 0.08\}$ for $\eta = 0.6$.    Single asterisks indicate that $a\to Z\gamma$ is kinematically allowed, while double asterisks indicate both $a\to Z\gamma$ and $a\to ZZ$ decay channels are possible.  

The exclusions for $\eta = 0.1$ and 0.3 in the bottom panel of Fig.~\ref{fig:GammacGammaspExclusion} are set by the same bin, while for $\eta =  0.6$ the peak of the photon spectrum has migrated far enough to higher energies that it is instead the neighboring bin that determines the exclusion. This results in the different shapes of the excluded regions obtained for $\eta = 0.6$ versus $\eta = 0.1$ and 0.3 that can be most easily seen in the green curves.  When we incorporate the above branching ratios into $Z\gamma$ and $ZZ$, the flux in the peak of the photon spectrum is reduced and the limits weaken accordingly, but the shapes of the curves remain the same.

\section{Summary and conclusions}
\label{sec:conclude}

We have performed a detailed study of DM annihilations in a hidden sector axion portal (HSAP) model where fermionic DM $\chi$ annihilates to pseudoscalars $a$.  We consider here  the regime where 
both the $p$-wave $\chi\chi\to aa$ and $s$-wave $\chi\chi\to aaa$ annihilation channels are kinematically available.  We find that the relic abundance is determined dominantly by $p$-wave annihilations at low DM masses, while $s$-wave contributions can dominate at higher DM masses.  In all cases the $s$-wave cross-section dominates the annihilation in the Milky Way today, and can be suppressed by as much as $10^{-4}$ relative to the standard expectation for a thermal WIMP.   
We take the pseudoscalar mediator $a$ to decay to SM states via an axion-like coupling to SM hypercharge  gauge bosons.  When $m_a<m_Z$, the pseudoscalar decays exclusively via $a\to \gamma\gamma$.  We derive the photon spectrum resulting from  $\chi\chi\to aaa\to6\gamma$ analytically in the limit of nonrelativistic DM annihilations.  This spectrum is dominated by a characteristic high-energy spectral feature that could help to facilitate discovery and identification of a DM signal in the busy environment of the Galactic Center.  When $m_a$ is large enough to allow for $Z\gamma$ and $ZZ$ decays, we determine the spectra for those decay channels numerically.  

DM density spikes around the Milky Way's supermassive black hole offer a speculative but irreplaceable potential discovery handle on this model.  
We estimated current sensitivity to HSAP DM annihilation within such spikes by comparing the predicted point-like gamma-ray spike signals to \textit{Fermi} and H.E.S.S. observations of Sgr A*.  We find interesting sensitivity to the HSAP model across a wide range of both particle and astrophysical parameters.  Canonical adiabatic spikes can probe cross-sections some four orders of magnitude below the standard thermal target, and are accordingly almost entirely ruled out in our model space.   However, substantial portions of parameter space remains open for shallower spikes,  particularly in less cuspy haloes.  In such (perhaps more realistic) astrophysical scenarios, we find that searches for spatially localized, sharp features in the gamma-ray spectrum remain well-motivated as a potential discovery handle on otherwise challenging models of DM.

The HSAP model compactly provides examples of both parametrically suppressed  annihilation cross-sections at low DM mass and mass-suppressed annihilation rates at high DM mass.   As is common for models of secluded DM \cite{Pospelov:2007mp},  the HSAP model also has severely suppressed signals at both direct detection and collider experiments, making indirect detection a vital experimental probe of this model.
This HSAP model is only one example of a rich variety of particle models that predict DM annihilation signals substantially below current observational sensitivity; the enhanced sensitivity to such faint annihilation signals in DM density spikes thus remains important to consider even as thermal targets are surpassed.   

\subsection*{Acknowledgments}
We are happy to acknowledge useful conversations with A. Srinivasan. We particularly thank J.~Tu for assistance with computing and B.~Mandava for assistance in algorithm efficiency improvement. The work of BTC was supported by the Lorella M.~Jones and the Philip J.~and Betty M.~Anthony Undergraduate Research Awards at the University of Illinois at Urbana-Champaign. The work of SLS is supported in part by NSF Grant PHY-1662211 and NASA Grant 80NSSC17K0070.  The work of JS is supported by DOE CAREER grant DE-SC0017840. 
\bibliography{MyBibTeX}
\bibliographystyle{JHEP}
\end{document}